\documentclass[twocolumn,amsmath,nofootinbib]{revtex4} 

\usepackage{url}
\usepackage{amssymb}
\usepackage{amsfonts}
\usepackage{amsbsy}
\usepackage{graphicx}
\usepackage{hyperref}
\usepackage{array} 
\usepackage{multirow}

\newcolumntype{x}[1]{%
>{\centering\hspace{0pt}}p{#1}}%
\providecommand{\openone}{\leavevmode\hbox{\small1\kern-3.8pt\normalsize1}}

\def\spose#1{\hbox to 0pt{#1\hss}}
\def\simlt{\mathrel{\spose{\lower 3pt\hbox{$\mathchar"218$}}
   \raise 2.0pt\hbox{$\mathchar"13C$}}}
\def\simgt{\mathrel{\spose{\lower 3pt\hbox{$\mathchar"218$}}
     \raise 2.0pt\hbox{$\mathchar"13E$}}}
 \def\simpropto{\mathrel{\spose{\lower 3pt\hbox{$\mathchar"218$}}
     \raise 2.0pt\hbox{$\propto$}}}

\def\beq#1{\begin{equation}\label{#1}}
\def\eeq{\end{equation}}
\def\beqa#1{\begin{eqnarray}\label{#1}}
\def\eeqa{\end{eqnarray}}

\def\ed{\end{document}}

\def\rn{}
\def\nn#1 #2{#2. #1}				
\def\nnn#1 #2 #3{#2. #3. #1}			
\def\nnnn#1 #2 #3 #4{#2. #3. #4 #1}		
\def\nnnnn#1 #2 #3 #4 #5{#2. #3. #4 #5. #1}	

\def\rf#1;#2;#3;#4;#5 {{\frenchspacing\par\rn#1, #3 {\bf #4}, #5 (#2). \par}}
\def\rg#1;#2;#3;#4;#5;#6 {{\frenchspacing\par\rn#1, #3 {\bf #4}, #5 (#2). \par}}
\def\rfbook#1;#2;#3;#4;#5 {{\frenchspacing\par\rn#1, {\it #3} (#5, #4, #2).\par}}
\def\rfprep#1;#2;#3 {{\par\frenchspacing\rn#1, #3 (#2).\par}}
\def\rfproc#1;#2;#3;#4;#5;#6 {{\frenchspacing\par\rn#1 #2, in {\it #3}, ed. #4 (#5: #6)\par}}
\def\rfprocp#1;#2;#3;#4;#5;#6;#7 {{\frenchspacing\par\rn#1 #2, in {\it #3}, ed. #4 (#5: #6), p#7\par}}

\begin{document}
\pdfoptionalwaysusepdfpagebox=5


\title{Consciousness as a State of Matter}

\author{Max Tegmark}

\address{Dept.~of Physics \& MIT Kavli Institute, Massachusetts Institute of Technology, Cambridge, MA 02139}

\date{Published in New Scientist, April 12, 2014 (pages 28-31).}

\vspace{10mm}

\begin{abstract}
I examine the hypothesis that consciousness can be understood as a state of matter, ``perceptronium", with distinctive information processing abilities. I explore five basic principles that may distinguish conscious matter from other physical systems such as solids, liquids and gases: the information, integration,  independence, dynamics  and utility principles.
This approach generalizes Giulio Tononi's integrated information framework for neural-network-based consciousness to arbitrary quantum systems,  and provides  interesting links to error-correcting codes and condensed matter criticality, as well as an interesting connections between the emergence of consciousness and the emergence of time.
(For more technical details, see \url{http://arxiv.org/abs/1401.1219}.)
\end{abstract}

\maketitle


Why are you conscious right now? Specifically, why are you having a subjective experience of reading these words, seeing colours and hearing sounds, while the inanimate objects around you presumably aren't having any subjective experience at all? 

Different people mean different things by ``consciousness", including awareness of environment or self. I am asking the more basic question of why you experience anything at all, which is the essence of what philosopher David Chalmers has coined ``the hard problem" of consciousness.

A traditional answer to this problem is dualism --- that living entities differ from inanimate ones because they contain some non-physical element such as an ``anima" or ``soul". Support for dualism among scientists has gradually dwindled. To understand why, consider that your body is made of about $10^{29}$ quarks and electrons, which as far as we can tell move according to simple physical laws. Imagine a future technology able to track all your particles: if they were found to obey the laws of physics exactly, then your purported soul is having no effect on your particles, so your conscious mind and its ability to control your movements would have nothing to do with a soul.

If your particles were instead found not to obey the known laws of physics because they were being pushed around by your soul, then we could treat the soul as just another physical entity able to exert forces on particles, and study what physical laws it obeys.

Let us therefore explore the other option, known as physicalism: that consciousness is a process that can occur in certain physical systems. This begs a fascinating question: why are some physical entities conscious, while others are not? If we consider the most general state of matter that experiences consciousness --- let's call it ``perceptronium" --- then what special properties does it have that we could in principle measure in a lab? What are these physical correlates of consciousness? Parts of your brain clearly have these properties right now, as well as while you were dreaming last night, but not while you were in deep sleep. 

Imagine all the food you have eaten in your life and consider that you are simply some of that food, rearranged. This shows that your consciousness isn't simply due to the atoms you ate, but depends on the complex patterns into which these atoms are arranged. If you can also imagine conscious entities, say aliens or future superintelligent robots, made out of different types of atoms then this suggests that consciousness is an emergent phenomenon (whose complex behaviour emerges from many simple interactions). In a similar spirit, generations of physicists and chemists have studied what happens when you group together vast numbers of atoms, finding that their collective behaviour depends on the patterns in which they are arranged: the key difference between a solid, a liquid and a gas lies not in the types of atoms, but in their arrangement. Boiling or freezing a liquid simply rearranges its atoms.

My hope is that we will ultimately be able to understand perceptronium as yet another state of matter. Just as there are many types of liquids, there are many types of consciousness. However, this should not preclude us from identifying, quantifying, modelling and understanding the characteristic properties that all liquid forms of matter, or all conscious forms of matter, share. Take waves, for example, which are substrate-independent in the sense that they can occur in all liquids, regardless of the liquid's atomic composition. Like consciousness, waves are an emergent phenomenon in the sense that they take on a life of their own: a wave can traverse a lake while the individual water molecules merely bob up and down, and the motion of the wave can be described by a mathematical equation that doesn't care what the wave is made of. 

Something analogous happens in computing: Alan Turing famously proved that all sufficiently advanced computers can simulate one another, so a video game character in her virtual world would have no way of knowing whether her computational substrate (``computronium") was a Mac or a PC, or what types of atoms the CPU was made of. All that would matter is abstract information processing. If this created character were complex enough to be conscious, like in the film The Matrix, then what properties would this information processing need to have?

I have long contended that consciousness is the way information feels when processed in certain complex ways. The neuroscientist Giulio Tononi has made this idea more specific and useful, making the compelling argument that for an information processing system to be conscious, its information must be integrated into a unified whole. In other words, it must be impossible to decompose the system into nearly independent parts --- otherwise these parts would feel like two separate conscious entities. Tononi and his collaborators have incorporated this idea into an elaborate mathematical formalism known as integrated information theory (IIT). 

IIT has generated significant interest in the neuroscience community, because it offers answers to many intriguing questions. For example, why do some information processing systems in our brains appear to be unconscious? Based on extensive research correlating brain measurements with subjectively reported experience, neuroscientist Christof Koch and others have concluded that the cerebellum --- a brain area whose roles include motor control --- is not conscious, but is an unconscious information processor that helps other parts of the brain with certain computational tasks. 

The IIT explanation for this is that the cerebellum is mainly a collection of ``feed-forward" neural networks in which information flows like water down a river, and each neuron affects mostly those downstream. If there is no feedback, there is no integration and hence no consciousness. The same would apply to Google's recent feed-forward artificial neural network that processed millions of YouTube video frames to determine whether they contained cats. In contrast, the brain systems linked to consciousness are strongly integrated, with all parts able to affect one another. 

IIT thus offers an answer to the question of whether a superintelligent computer would be conscious: it depends. A part of its information processing system that is highly integrated will indeed be conscious. However, IIT research has shown that for many integrated systems, one can design a functionally equivalent feed-forward system that will be unconscious. This means that so-called ``p-zombies" can, in principle, exist: systems that behave like a human and pass the Turing test for machine intelligence, yet lack any conscious experience whatsoever. Many current ``deep learning" AI systems are of this p-zombie type. Fortunately, integrated systems such as those in our brains typically require much fewer computational resources than their feed-forward ``zombie" equivalents, which may explain why evolution has favoured them and made us conscious.

Another question answered by IIT is why we are unconscious during seizures, sedation and deep sleep, but not REM sleep. Although our neurons remain alive and well during sedation and deep sleep, their interactions are weakened in a way that reduces integration and hence consciousness. During a seizure, the interactions instead get so strong that vast numbers of neurons start imitating one another, losing their ability to contribute independent information, which is another key requirement for consciousness according to IIT. This is analogous to a computer hard drive where the bits that encode information are forced to be either all zeros or all ones, resulting in the drive storing only a single bit of information. Tononi, together with Adenauer Casali, Marcello Massimini and other collaborators, recently validated these ideas with lab experiments. They defined a ``consciousness index" that they could measure by using an EEG to monitor the brain's electrical activity after magnetic stimulation, and used it to successfully predict whether patients were conscious.

Awake and dreaming patients had comparably high consciousness indices, whereas those anaesthetised or in deep sleep had much lower values. The index even successfully identified as conscious two patients with locked-in syndrome, who were aware and awake but prevented by paralysis from speaking or moving . This illustrates the promise of this technique for helping doctors determine whether unresponsive patients are conscious. 

Despite these successes, IIT leaves many questions unanswered. If it is to extend our consciousness-detection ability to animals, computers and arbitrary physical systems, then we need to ground its principles in fundamental physics. IIT takes information, measured in bits, as a starting point. But when I view a brain or computer through my physicist's eyes, as myriad moving particles, then what physical properties of the system should be interpreted as logical bits of information? I interpret as a ``bit" both the position of certain electrons in my computer's RAM memory (determining whether the micro-capacitor is charged) and the position of certain sodium ions in your brain (determining whether a neuron is firing), but on the basis of what principle? Surely there should be some way of identifying consciousness from the particle motions alone, even without this information interpretation? If so, what aspects of the behaviour of particles corresponds to conscious integrated information?

The problem of identifying consciousness in an arbitrary collection of moving particles is similar to the simpler problem of identifying objects there. For instance, when you drink iced water, you perceive an ice cube in your glass as a separate object because its parts are more strongly connected to one another than to their environment. In other words, the ice cube is both fairly integrated and fairly independent of the liquid in the glass. The same can be said about the ice cube's constituents, from water molecules all the way down to atoms, protons, neutrons, electrons and quarks. Zooming out, you similarly perceive the macroscopic world as a dynamic hierarchy of objects that are strongly integrated and relatively independent, all the way up to planets, solar systems and galaxies.

This grouping of particles into objects reflects how they are stuck together, which can be quantified by the amount of energy needed to pull them apart. But we can also reinterpret this in terms of information: if you know the position of one of the atoms in the piston of an engine, then this gives you information about the whereabouts of all the other atoms in the piston, because they all move together as a single object. A key difference between inanimate and conscious objects is that for the latter, too much integration is a bad thing: the piston atoms act much like neurons during a seizure, slavishly tracking one another so that very few bits of independent information exist in this system. A conscious system must thus strike a balance between too little integration (such as a liquid with atoms moving fairly independently) and too much integration (such as a solid), suggesting that consciousness is maximised near a phase transition between less- and more-ordered states; indeed, humans lose consciousness unless key physical parameters of our brain are kept within a narrow range of values. 

An elegant balance between information and integration can be achieved using error-correcting codes: methods for storing bits of information that know about each other, so that all information can be recovered from a fraction of the bits. These are widely used in telecommunications, as well as in the ubiquitous QR codes from whose characteristic pattern of black and white squares your smartphone can read a web address. As error correction has proven so useful in our technology, it would be interesting to search for error-correcting codes in the brain, in case evolution has independently discovered their utility --- and perhaps made us conscious as a side effect. 

We know that our brains have some ability to correct errors because you can recall the correct lyrics for a song you know from a slightly incorrect fragment of it. John Hopfield, a biophysicist renowned for his eponymous neural network model of the brain, proved that his model has precisely this error-correcting property. However, if the hundred billion neurons in our brain do form a Hopfield network, calculations show that it could only support about 37 bits of integrated information --- the equivalent of a few words of text. This begs the question of why the information content of our conscious experience seems to be significantly larger than 37 bits. The plot thickens when we view our brain's moving particles as a quantum-mechanical system. As I showed in January, the maximum amount of integrated information then drops from 37 bits to about 0.25 bits, and making the system larger doesn't help (\url{arxiv.org/abs/1401.1219}). 

This integration problem can be circumvented by adding another principle to the list that a physical system must obey in order to be conscious. So far I have outlined three: the information principle (it must have substantial information storage capacity), the independence principle (if must have substantial independence from the rest of the world) and the integration principle (it cannot consist of nearly independent parts). The aforementioned 0.25 bit problem can be bypassed if we also add the dynamics principle --- that a conscious system must have substantial information-processing capacity, and it is this processing rather than the static information that must be integrated. For example, two separate computers or brains can't form a single consciousness.  

These principles are intended as necessary but not sufficient conditions for consciousness, much like low compressibility is a necessary but not sufficient condition for being a liquid. As I explore in my book Our Mathematical Universe, this leads to promising prospects for grounding consciousness and IIT in fundamental physics, although much work remains and the jury is still out on whether it will succeed.

If it does succeed, this will be important not only for neuroscience and psychology, but also for fundamental physics, where many of our most glaring problems reflect our confusion about how to treat consciousness. In Einstein's theory of general relativity, we model the ``observer" as a fictitious disembodied massless entity having no effect whatsoever on that which is observed. In contrast, the textbook interpretation of quantum mechanics states that the observer does affect the observed. Yet after a century of spirited debate, there is still no consensus on how exactly to think of the quantum observer. Some recent papers have argued that the observer is the key to understanding other fundamental physics mysteries, such as why our universe appears so orderly, why time seems to have a preferred forward direction, and even why time appears to flow at all. 

If we can figure out how to identify conscious observers in any physical system and calculate how they will perceive their world, then this might answer these vexing questions.

\end{document}